\documentclass[cameraready]{Interspeech}

\title{Utterance-Level Methods for Identifying Reliable ASR-Output for Child Speech}

\author[affiliation={12}]{Gus}{Lathouwers}
\author[affiliation={12}]{Lingyun}{Gao}
\author[affiliation={12}]{Catia}{Cucchiarini}
\author[affiliation={12}]{Helmer}{Strik}

\address{
    $^1$ Centre for Language Studies, Radboud University, Netherlands,\\ 
    $^2$ Department of Language and Communication, Radboud University, Netherlands
}

\email{guslathouwers@gmail.com, lingyun.gao@ru.nl, catia.cucchiarini@ru.nl, helmer.strik@ru.nl}


\usepackage{comment}
\usepackage{tabularx}

\begin{document}

\maketitle

\begin{abstract}
    Automatic Speech Recognition (ASR) is increasingly used in applications involving child speech, such as language learning and literacy acquisition. However, the effectiveness of such applications is limited by high ASR error rates. The negative effects can be mitigated by identifying in advance which ASR-outputs are reliable. This work aims to develop two novel approaches for selecting reliable ASR-output at the utterance level, one for selecting reliable read speech and one for dialogue speech material. Evaluations were done on an English and a Dutch dataset, each with a baseline and finetuned model. The results show that utterance-level selection methods for identifying reliably transcribed speech recordings have high precision for the best strategy (P \textgreater \ 97.4) for both read speech and dialogue material, for both languages. Using the current optimal strategy allows 21.0\% to 55.9\% of dialogue/read speech datasets to be automatically selected with low (UER of \textless \ 2.6) error rates.

\end{abstract}

\section{Introduction}

Automatic Speech Recognition (ASR) technology has been developed for and applied to different downstream tasks in speech processing, such as language-learning and learning-to-read software \cite{Russell2024,VanDoremalen2016,Bai2021}. ASR may specifically be employed in children speech as an important component in tools that support speech diagnostics, language learning, or learning to read in classroom environments \cite{Bhardwaj2022,Shadiev2023,Bai2025}. In this context, there is a growing interest in how ASR can be utilized to reduce human effort, for example reducing the time teachers spend manually transcribing child speech for mandatory language skill tests in elementary school \cite{Groenhof2025,Harmsen2025}. 

Despite the interest in employing ASR in such usecases and its potential benefits, the application of ASR to child speech is hampered by a number of factors, including limited availability of child speech data to train ASR models \cite{Jain2023}, variable child speaker characteristics \cite{Yadav2021}, high amounts of noise in child speech \cite{Dutta2022}, and ASR models typically not trained for transcribing verbatim speech \cite{Russell2024}. Though methods such as finetuning existing ASR models on child speech \cite{Shekoufandeh2025} and data-augmentation \cite{Zhang2024} can increase child speech ASR accuracy, as a whole performance still lags behind that of models applied to adult speech  \cite{Shivakumar2022}.

In addition to reducing recognition error rates in child speech, an adjoining task concerns identifying the cases in which ASR transcribes speech reliably. For example, high noise in classroom settings detrimentally affects ASR accuracy \cite{Dutta2022}, and similarly the presence of pronunciation errors in speech has been associated with higher numbers of errors in subsequent ASR-transcriptions \cite{Alderete2025}. Given that current state-of-the-art child speech models may still suffer from error rates anywhere between 5\% to 50\% on the word level \cite{Zhang2024,Shekoufandeh2025}, being able to recognize the scenarios in which ASR is successful can increase the effectiveness of tools that make use of ASR.

Traditionally, methods to measure the reliablity of ASR-output have relied on confidence estimation, which has a long-standing research history \cite{Sukkar1996,Li2021} and employs different computational techniques to generate confidence scores \cite{Li2021,Oneata2021}. However, confidence estimation techniques may be unreliable in certain cases, especially in scenarios with high degrees of uncertainty or ambiguous speech patterns in the source audio \cite{Kuhn2025}, which is often the case in child speech \cite{Rumberg2023}. A related field of investigation concerns quality estimation. Like confidence estimation, quality estimation aims to assess the veracity of ASR-transcriptions, but methods typically focus on the quality of output strings generated by ASR-models, instead of relying on internal model probabilities observed during the generation of ASR-output \cite{Negri2014}. Quality estimation methods often operate at utterance level, analyzing linguistic coherence \cite{Jalalvand2018,Javadi2024}, and thus do not require access to any internal model states like confidence estimation methods do.

Research on error detection and remediation in ASR-output has focused on post-processing techniques, such as pipelines using Large Language Models (LLMs) to improve transcription quality \cite{Gao2025a}. LLMs may also be effective at correcting output based on semantic properties within text, such as using semantic context to correct user queries in ASR-systems \cite{Asano2025}. For instance, sentences that contain semantic outliers such as nonsensical words may signal the ASR made errors during the transcription process. Within utterances or phrases, semantic information has been used to score individual words and score ASR-hypotheses accordingly \cite{Liu2019}. 

\begin{figure*}[t]
  \centering
  \includegraphics[width=\textwidth]{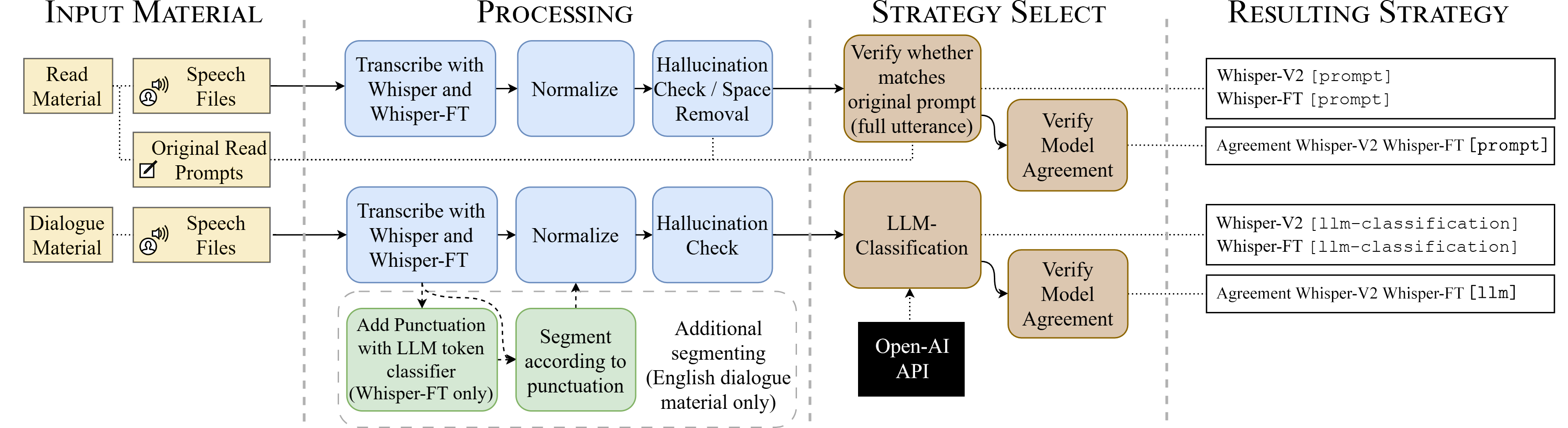}
  \caption{Audio Processing Pipeline for Different Strategies used.}
  \label{fig:english_diagram}
\end{figure*}

In summary, existing ASR reliability confidence estimation methods may struggle when applied to child speech, which typically contains noise and is difficult to transcribe \cite{Kuhn2025,Rumberg2023}. The aim of this research is to develop a novel approach to assess the reliability of ASR-output in child speech settings within individual utterances. Previously, semantic context has been used to score ASR-hypotheses, such as for conversational AI usage \cite{Asano2025}. Here, we draw on concepts associated with semantic context and grammatical consistency within phrases as a means to determine the reliability of child speech utterances transcriptions made by ASR. The usecases of such methods lie in (a) automatically identifying parts of data speech that are likely to be correctly transcribed by an ASR-model, and thus do not require human verification in the case of newly generated data, and (b) automatic detection of parts of correctly transcribed dialogue speech by ASR-models, which can be used to select high-quality ASR-output automatically. To the best of our knowledge, the novelty of this work lies in assessing the reliability of ASR-output using newly developed methods on an utterance level specifically for read and dialogue material, which differ from traditional confidence estimation methods that typically function at the word level \cite{Negri2014}. The research is also novel for including a method for selecting reliable ASR-transcripts for longer speech recordings, such as in the case of dialogue material. The research question (RQ) we address is as follows:

\begin{itemize}
\item RQ: To what extent are utterance-level classification methods effective at identifying reliable ASR output in read and dialogue speech material? 
\end{itemize}

\section{Methodology}

In the current study, we select reliable output at the utterance level. Two methods for selecting reliable ASR-output were tested across two types of materials, namely one for read speech which makes use of the original read prompt commonly present in read material, and one for dialogue audio that can be applied when no original read prompt is present. Dialogue and read material is tested across two languages, Dutch and English\footnote{See  \texttt{http://github.com/anonimized} for the full pipeline}.

\subsection{Datasets}

To address our research question, two datasets were used: JASMIN for Dutch and CSLU for English. For JASMIN \cite{cucchiarini2008recording}, all  speech recordings from Dutch native children aged 7-11 were used. The total length of the dataset is 9 hours and 51 minutes, distributed across 10,642 utterances (M=3.34, SD=2.35). The majority of JASMIN (71.9\%) comprises read speech material and the remainder dialogue material (28.1\%). For all reading material, the original prompt the child was instructed to read was included in the dataset. For both reading and dialogue material, a manual annotation made by a human transcriber was included in the data set. For JASMIN, 80\% was allocated for model fine-tuning, while the remaining 20\% was allocated for evaluation (see Section~\ref{subsec:asr_models}). The evaluation set was n=2,129 in size, with n=1,551 read material utterances and n=578 dialogues. For the read material, a little more than half of the utterances (54.4\%) were incorrectly read by the children.

For English, CSLU \cite{shobaki2000} was used. Because the original dataset is over 100 hours in length, a subset totalling 5 hours in length (3,534 utterances, M=4.94, SD=12.60) was randomly sampled from the total dataset , with only utterances from children in grades 2 to 6 included. This was done to mirror the age group 
of the JASMIN dataset, ages 7-11, and thus allow comparability. As with JASMIN, the majority of the utterances are read material (70.4\%), and the remainder dialogues. In the reading material, each individual utterance had the original prompt included, as well as a classification made by a human transcriber on whether the prompt was read correctly by the child (true = correct, false = incorrect). For read utterances, 6.7\% were classified as being incorrectly pronounced. For the dialogue data, unlike JASMIN, child phrases were not segmented into short utterances, but instead were full-length recordings ranging from 11 to 479 seconds (M=100.1 sd=34.5). Because evaluations in the current study are conducted at the utterance level, additional post-processing was required to segment the longer English dialogues into individual utterances (see Section~\ref{subsec:eng_dialogue}).

\subsection{ASR Models}
\label{subsec:asr_models}

For both English and Dutch, the base non-modified version of Whisper (Whisper-V2, 1550 million parameters) as well as a finetuned medium version (Whisper-FT, 769 million parameters) were evaluated. The standard version was applied to both Dutch and English and served as a baseline, whereas the finetuned version was language-specific. For English, a model made public by \cite{Jain2023} was referenced. This particular model was chosen because it generalizes well to a number of English child datasets in the study, and because it was trained on a roughly equal number of hours (PF-STAR, 10 hours) as the Dutch finetuned model. For Dutch, because there was no publicly available finetuned model, a medium version of finetuned was trained on a partial allocation of the JASMIN dataset. In accordance with another study that used the same dataset \cite{Shekoufandeh2025}, training was halted after a 5-epoch schedule, with a learning rate set 1e-5. Fine-tuning took approximately 28 hours on a single RTX A6000. All fine-tuning was conducted using the command prompt with the Hugging Face and Torch python libraries.

\begin{table*}[t]
  \caption{Performance Analysis of Methods for Selecting Reliable ASR-Output}
  \label{tab:mainresults}
  \centering
  \begin{tabular}{l c c c c c c c c}
    \toprule

    \multicolumn{1}{l}{\textbf{}} &
    \multicolumn{4}{c}{\textbf{Dutch}} &
    \multicolumn{4}{c}{\textbf{English}} \\

    \cmidrule(rl){2-5} \cmidrule(rl){6-9}

    \textbf{Condition} &
    \textbf{P} &
    \textbf{R} &
    \textbf{F1} &
    \textbf{MCC} &
    \textbf{P} &
    \textbf{R} &
    \textbf{F1} &
    \textbf{MCC} \\

    \midrule
    \multicolumn{9}{l}{\textsc{Read material}} \\
    \hspace{0.2cm}Whisper-V2 \texttt{[prompt]}                        
    & 80.7 & 60.8 & 69.3 & 0.51 & 97.8 & 74.0 & 84.3 & 0.28
    \\

    \hspace{0.2cm}Whisper-FT \texttt{[prompt]}                       & 97.2 & 91.5 & 94.3 & 0.90 & 98.2 & 69.3 & 81.3 & 0.27
    \\
       
    \hspace{0.2cm}Agreement Whisper-V2 and Whisper-FT \texttt{[prompt]}                     & 98.3 & 57.9 & 72.9 & 0.64 & 98.4 & 79.5 & 88.0 & 0.71   
    \\

    \midrule
    \multicolumn{9}{l}{\textsc{Dialogue material}} \\
    \hspace{0.2cm}Whisper-V2 \texttt{[LLM-classification]}
    & 62.1 & 95.6 & 75.3 & 0.48 & 86.0 & 56.6 & 68.3 & 0.28         
    \\
    
    \hspace{0.2cm}Whisper-FT \texttt{[LLM-classification]}
    & 88.9 & 93.8 & 91.3 & 0.62 & 83.4 & 74.9 & 79.0 & 0.54            
    \\
    
    \hspace{0.2cm}Agreement Whisper-V2 and Whisper-FT \texttt{[llm-classification]}
    & 97.4 & 53.8 & 69.3 & 0.45 & 98.0 & 28.0 & 43.6 & 0.29 
    \\

    \bottomrule
  \end{tabular}

\end{table*}

\subsection{Metrics}

In the current study, selecting reliable output is done at the
utterance level. For the reading material, utterances may comprise single individual words or short sentences. In the dialogue material, utterances are always sentences. Precision (P), Recall (R), F1 score (F1), and Mathews Correlation Coefficient (MCC) were calculated to determine how effect methods are at selecting reliable ASR output. MCC was included due to the advantage of being a reliable statistic in the case of class imbalances \cite{Chicco2023}.

For selecting reliable outputs, two main methods were developed that to the best of our knowledge have not been used elsewhere: \texttt{[prompt]} for the read material, \texttt{[LLM-classification]} for the dialogue material. The testing procedure is as follows (see Figure~\ref{fig:english_diagram}). First, an ASR-model is applied to a number of utterances across a dataset. Standard text normalization is applied to all ASR-outputs, as well as one to two additional functions removing some hallucinated output and detecting erroneous space insertions. Then, one of the two methods of selecting reliable output is applied, namely \texttt{[prompt]} for read data and \texttt{[LLM-classification]} for dialogue data. This results in a prediction for each utterance, namely positive or negative. Here, "positive" indicates that the ASR-output is classified as being reliable, with "negative" indicating non reliable output. Afterwards, it is verified whether this prediction of positive or negative is correct by contrasting it with the manually verified solution (true or false) found in the datasets. In the current study, given that our aim is to identify reliable output, any method that maximizes true positives while keeping false positives at a minimum constant is considered effective. 

In addition to calculating P, R, F1, and MCC, the total size of the full dataset isolated by each method is also displayed in a separate table (Table~\ref{tab:subsetresults}). This represents the subsection out of the full dataset a method had identified as being reliable, and the final proportion of the full dataset the method would mark automatically with an accompanying error rate (expressed in Word Error Rate, WER, and Utterance Error Rate, UER). Given that the current work aims at identifying at the utterance level, UER serves as the primary benchmark. UER is defined as the percentage of utterances out of the full dataset that are selected as reliable (positive) but contain transcription errors. Listed below are descriptions of the two methods used to identify reliable output. For read material, \texttt{[prompt]} was used. For dialogue material, given that there was no original prompt available for this material type, \texttt{[LLM-classification]} was used. Also, for each data type (read, dialogue), an additional strategy was tested that only treats utterances as positive if both models (Whisper-v2 and Whisper-FT) used within a language produce the same ASR-output for an utterance.

\subsubsection{Checking against prompt}

As noted earlier, for each utterance in the reading material in the English and Dutch datasets, the prompt the child originally read was included. The proposed method here is to check an ASR output against this originally read prompt as an indicator of whether the ASR output is grammatically and linguistically sound and therefor likely to be reliable. The process is as follows: an ASR model is applied to a speech utterance, resulting in an ASR output (AO), for which then it is checked whether the AO fully matches at the utterance level the originally intended prompt (PR) the child had to read. If this is the case, the utterance is classified as a positive, if not as a negative. For evaluation, it is checked whether the positive is true by comparing  it against the manual annotation (MO).

\subsubsection{LLM-classification}

Although there can be mismatches, read speech recordings generally contain the originally read prompts for the utterances; however, in extemporaneous dialogues this is not the case. Thus, a different approach needs to be taken to detect semantic or grammatical oddities. In the current research, this is done by having an LLM classify whether an ASR output is likely reliable or not, based on the following instructions\footnote{See Github repository for Dutch/English-tailored instructions.}:

\begin{quote}
\textit{"The following sentence is a transcription of a child utterance. Indicate whether it contains repeated words, accidentally pronounced small vowels in between, or if there are other oddities present, such as strange words that do not fit:} \texttt{<sentence>}.  \textit{Answer with 'wrong' if this is the case or 'correct' otherwise."}
\end{quote}

Here, \texttt{<sentence>} is replaced by the ASR-output of a particular utterance. If the LLM classification starts with 'correct,' it is considered a positive. Any LLM classification starting with 'wrong' is considered a negative. For all LLM classifications, ChatGPT-5 was selected as the model with reasoning effort and verbosity set to 'low.' All classifications were gathered using the OpenAI API (publicly available snapshot \texttt{gpt-5-2025-08-07}). 

\subsubsection{Model agreement as an additional filter}

Because each language had two ASR-models, an additional filter for identifying output reliability checks whether the baseline and finetuned versions generate identical output for a particular utterance. Therefore, for this strategy, an additional condition is that the two models give matching output for it to be considered a positive. Thus, for read material agreement, a positive here is if \textit{both} models give output for a particular utterance that match the prompt. Similarly, for dialogue material, model agreement strategy implies that \textit{both} models output identical text for utterances, and that the LLM-classification classifies these as reliable.

\begin{table*}[t]
  \caption{Subportion of Datasets Selected with Accompanying UER and WER}
  \label{tab:subsetresults}
  \centering
  \begin{tabular}{l c c c c c c}
    \toprule

    \multicolumn{1}{l}{\textbf{}} &
    \multicolumn{3}{c}{\textbf{Dutch}} &
    \multicolumn{3}{c}{\textbf{English}} \\

    \cmidrule(rl){2-4} \cmidrule(rl){5-7}

    \textbf{Condition} &
    \textbf{of Dataset} &
    \textbf{UER} &
    \textbf{WER} &
    \textbf{of Dataset} &
    \textbf{UER} &
    \textbf{WER}$^{\dagger}$ \\

    \midrule
    \multicolumn{7}{l}{\textsc{Read material}} \\
    
    \hspace{0.2cm}Whisper-V2                        
    & 100.0\% & 56.2 & 16.7 & 100.0\% & 29.4 & -
    \\

    \hspace{0.2cm}Whisper-V2 \texttt{[prompt]}                        
    & 34.0\% & 19.3 & 5.2 & 70.6\% & 2.2 & -
    \\

    \hspace{0.2cm}Whisper-FT                      
    & 100.0\% & 24.3 & 5.1 & 100.0\% & 34.1 & -
    \\        

    \hspace{0.2cm}Whisper-FT\texttt{[prompt]}                       
    & 42.1\% & 2.8 & 0.5 & 65.9\% & 1.8 & -
    \\    
       
    \hspace{0.2cm}Agreement Whisper-V2 and Whisper-FT \texttt{[prompt]}                    & 26.6\% & 1.7 & 0.3 & 55.9\% & 1.6 & -
    \\

    \midrule
    \multicolumn{7}{l}{\textsc{Dialogue material}} \\
    \hspace{0.2cm}Whisper-V2 
    & 100.0\% & 52.6 & 37.8 & 100.0\% & 32.0 & 6.8 
    \\

    \hspace{0.2cm}Whisper-V2 \texttt{[LLM-classification]}
    & 73.0\% & 37.9 & 28.8 & 48.0\% & 14.0 & 3.4 
    \\    
    
    \hspace{0.2cm}Whisper-FT
    & 100.0\% & 24.7 & 14.4 & 100.0\% & 42.8 & 10.8 
    \\

    \hspace{0.2cm}Whisper-FT \texttt{[LLM-classification]}
    & 79.4\% & 11.1 & 5.8 & 51.4\% & 16.6 & 3.5              
    \\    
    
    \hspace{0.2cm}Agreement Whisper-V2 and Whisper-FT \texttt{[llm-classification]}
    & 40.5\% & 2.6 & 1.5 & 21.0\% & 2.0 & 0.6
    \\

    \bottomrule
  \end{tabular}
\begin{flushleft}
\footnotesize{$^{\dagger}$WER is not reported for English read material because no manual annotations were available for this material type for English.}
\end{flushleft}
\end{table*}

\subsection{English dialogue processing}
\label{subsec:eng_dialogue}

For the English dialogue data, the process of identifying reliable ASR-output was more complicated because the CSLU English dialogue data were recorded as speech files up to 3 minutes in length. To make predictions at an utterance-level, the full output needed to be segmented. The process was as follows (see Figure~\ref{fig:english_diagram}). Both Whisper-V2 and Whisper-FT were exposed to the original speech recordings and provided full length output. ASR-Output for Whisper-V2 contained punctuation, any \texttt{','} or \texttt{'.'} was used to segment the transcription into separate utterances. For the English Whisper-FT, the ASR-model generated no punctuation. For this model, punctuation was added using a token classifier that adds punctuation to any text. Afterwards, as with Whisper-V2 output, the full transcript was segmented at all \texttt{','} or \texttt{'.'} points. Finally, for both Whisper-V2 and Whisper-FT output, word-alignment with the correct manual annotation was performed using the Python package \texttt{jiwer} to establish which utterances were error-free. Because Whisper-V2 and Whisper-FT had different transcriptions with punctuation at different positions, the final number of utterances after segmentation was different (n=729 for Whisper-V2, n=774). However, model agreement could still easily be assessed by cross-checking lists of segmented utterances.

\section{Results}

Table~\ref{tab:mainresults} displays precisons for all three strategies read material (top 4 rows) and dialogue material (bottom 4 rows). For both material types, for both languages, the highest P values are obtained for the model agreement conditions, with P values higher than 97.3. For read material only, all model/language combinations achieved  P \textgreater \ 97.1, with the exception of the Dutch Whisper-V2 model, which achieved a P of 80.7. None of the individual models for dialogue material achieved P of \textgreater \ than 88.9.

Table~\ref{tab:subsetresults} displays the percentages of subset selected by each strategy, and the accompanying error rates on an utterance (UER) and word (WER) level. It can be observed that the error rates for the subsets selected by applying different strategies (i.e. \texttt{[prompt]}, \texttt{[llm-classification]}) are lower, and that the best results are once again obtained for the combination agreement models, with error rates smaller than 2\% for read speech, and smaller than 3\% for dialogue speech at the utterance level. Note that the UER values for the subsets in Table~\ref{tab:subsetresults}  are the compliments of the P values in Table~\ref{tab:mainresults}  (P + UER = 100, i.e. for Dutch: 80.7 + 19.3, 97.2 + 2.8, 98.3 + 1.7; etc.).

\section{Discussion and Conclusion}

In the current study, for read material, for both Dutch and English, using just the finetuned model to select utterances resulted in high precision (P \textgreater \ 97) across respectively 42.1\% (Dutch) and 65.9\% (English) of the full read material datasets. The higher number of cases selected in the English dataset is likely attributable to the lower number of utterances that contained errors by children in reading the prompts in comparison with the Dutch dataset used here. The strategy leveraging the two different models resulted in high precision (P \textgreater \ 97.4) across all languages and material types, selecting between 21.0\% to 55.9\% of all data depending on the language and material type.

The techniques presented here can identify good portions of full datasets, partially bypassing the need for manual annotation \cite{Widodo2014}. Similarly, because the optimal strategy achieved high precision for long recordings of dialogue data, it could not only potentially be used in other material types such as child stories \cite{Bai2021} but also in more general speech software tools \cite{Russell2024}. Since traditional confidence estimation techniques may struggle especially in scenarios with high degrees of noise \cite{Kuhn2025}, the current methods may complement these, as they show high accuracy even with elevated error rates for baseline models. 

The current study presented two methods for identifying reliable ASR output for child speech, including long speech files. These methods can be used in utterance level comparisons with good results for the best strategy. Limitations are that the proposed methods cannot be used to detect ASR-output that is correct, but contains grammatical or semantic errors, and that the method used for selecting reliable dialogue can only select sentences. Future research could focus on applying the proposed methods to other languages or material types, combining these with reliability estimation techniques such as confidence estimation, and further test to see how well they generalize to other age groups such as adult speech.

\section{Generative AI Use Disclosure}

No generative AI was used for generating core content. Generative AI tools were only used for editing and polishing the language and checking the grammar of this manuscript to improve clarity and readability. The authors of the current paper are fully responsible for its contents.

\bibliographystyle{IEEEtran}
\bibliography{mybib}

\end{document}